\documentclass[a4paper]{article}
\usepackage{indentfirst}
\usepackage{hyperref}
\usepackage{multirow}
\usepackage{latexsym,bm,amsmath,amssymb}
\usepackage{graphicx}
\usepackage{epsfig}
\usepackage{subfigure}
\usepackage{cite}
\usepackage{color}
\usepackage[top=1in, bottom=1in, left=1.25in, right=1.25in]{geometry}
\usepackage{slashed}
\usepackage{fancyhdr}
\usepackage{slashed}
\usepackage{makecell,rotating,multirow,diagbox}
\usepackage{tikz}
\usepackage{float}
\usepackage{appendix}
\usepackage{setspace}
\usepackage{geometry}
\usepackage{graphics}
\usepackage{multirow}
\usetikzlibrary{shapes.geometric, arrows}
\usetikzlibrary{trees}
\usetikzlibrary{decorations.pathmorphing}
\usetikzlibrary{decorations.markings}
\tikzset{
        photon/.style={decorate, decoration={snake}, draw=red},
        nucleon/.style={draw=black, postaction={decorate},
           decoration={markings,mark=at position .55 with{\arrow[draw=black]{>}}}},
        pion/.style={draw=blue, postaction={decorate},
        decoration={markings,mark=at position .55 with{\arrow[draw=blue]{}}}},
        sigma/.style={draw=black, postaction={decorate},
        decoration={markings,mark=at position .55 with{\arrow[draw=black]{}}}},
        link/.style    = { draw=black, double = white, line width = 1.8pt, double distance = 0.8pt , postaction={decorate},decoration={markings,mark=at position .55 with{\arrow[draw=black]{>}}}},
    }

\renewcommand{\[}{\left [}
\renewcommand{\]}{\right ]}

\newcommand{\be}{\begin{equation}}
\newcommand{\ee}{\end{equation}}
\newcommand{\ba}{\begin{array}{c}} \newcommand{\ea}{\end{array}}
\newcommand{\bqa}{\begin{eqnarray}}
\newcommand{\eqa}{\end{eqnarray}} 

\newcommand{\no}{\nonumber\\}

\def\bea{\arraycolsep .1em \begin{eqnarray}}
\def\eea{\end{eqnarray}}

\def\s0#1#2{\mbox{\small{$ \frac{#1}{#2} $}}}
\def\0#1#2{\frac{#1}{#2}}

\begin{document}

\setcounter{topnumber}{10}
\setcounter{totalnumber}{50}
\title{An $N/D$ study  of the $S_{11}$ channel $\pi N$ scattering amplitude   }
\maketitle

\begin{center}
{\sc
Qu-Zhi Li,$^{\dagger\,}$ \,
Yao Ma,$^{\dagger\,}$ \,
Wen-Qi~Niu,$^{\dagger\,}$ \,
Yu-Fei~Wang,$^\ddagger$ \,
Han-Qing~Zheng$^{\heartsuit\,,\star\,,}$\footnote{Corresponding author.}}
\\
\vspace{0.5cm}
\noindent{\small{$^\dagger$ \it  Department of Physics and State Key
Laboratory of Nuclear Physics and Technology,
 Peking University, Beijing 100871, P.~R.~China}}\\
\noindent{\small{$^\ddagger$ {\it  Institute for Advanced Simulation, Institut f{\"u}r Kernphysik and J\"ulich Center for Hadron Physics, Forschungszentrum J{\"u}lich, D-52425 J{\"u}lich, Germany}}}\\
\noindent{\small{$^\heartsuit$ \it  College of Physics, Sichuan University, Chengdu, Sichuan 610065, P.~R.~China}}\\
\noindent{\small{$^\star$ \it   Collaborative Innovation Center of
Quantum Matter, Beijing, Peoples Republic of China}}
\end{center}
\begin{abstract}
Extensive dynamical $N/D$ calculations are made in the study of $S_{11}$ channel low energy $\pi$N scatterings, based on various phenomenological model inputs of left cuts at tree level. The subtleties of the singular behavior of the partial wave amplitude at the origin of the complex $s$ plane are carefully analysed. Furthermore, { it is  found that the dispersion representation for the phase shift, $\delta$, has to be modified in the case of $\pi $N scatterings. An additional contribution from the dispersion integral exists, which is, however, almost exactly cancelled the contribution from two virtual poles located near the end points of the segment cut induced by $u$ channel nucleon exchanges.} Relying very little on the details of the dynamical inputs, the subthreshold resonance $N^*(890)$ survives.
\end{abstract}
\vspace{1cm}
\section{Introduction}\label{intro}
In a series of recent publications~\cite{Wang:2018nwi}\cite{Wang:2018gul}\cite{Wang:2017agd}, it is suggested that there exists a sub-threshold $1/2^-$ nucleon resonance hidden in $S_{11}$ channel of $\pi N$ scatterings, with a pole mass $\sqrt{s}=(0.895 \pm 0.081) -(0.164 \pm 0.023)i$ GeV. The result is obtained by using the production representation (PKU representation) for partial wave amplitudes~\cite{Zheng:2003rw}\cite{Zhou:2006wm}\cite{Zhou:2004ms}\cite{Xiao:2000kx}\cite{He:2002ut}.
 It is found later that the $N^*(890)$ pole may also be generated from a conventional and simple $K$-matrix fit, though the latter suffers from the existence of spurious poles on 1st Riemann sheet of complex $s$ plane~\cite{Ma:2020sym}. Properties of $N^*(890)$ are also investigated, such as its coupling to {$N\gamma$ and $N\pi$~\cite{Ma:2020hpe, Cao:2021kvs}.} It is found that its coupling to $N\pi$ is considerably  larger than that of the $N^*(1535)$, while its coupling to $N\gamma$ is comparable to that of the $N^*(1535)$. These results on couplings look reasonable and are within expectations, hence providing further evidence on the existence of $N^*(890)$.

However, to firmly establish the existence of such a subthreshold resonance is still a difficult task. Besides dispersion relations, the most frequently used tools at hand  are perturbation chiral amplitudes and their unitarizations (For a recent review, see Ref.~\cite{Yao:2020bxx}), or (unitarized) resonance models. However, these unitarization techniques are far from being perfect when used in the study of low energy strong interaction physics. Especially, when applying to partial wave amplitudes with unequal mass scatterings, extra difficulties will occur, as will be discussed at some lengths in this paper. The  major difficulties arose at $s=0$ point in $s$ plane, where chiral expansions break down since chiral expansions and partial wave projections do not commute when $s\to 0$. The expected decoupling property of heavy  resonances when their masses sent to infinity is also violated in partial wave amplitudes at $s=0$, for a purely kinematical reason in partial wave projections with unequal mass scatterings. The major task of this paper is to show how the subthreshold resonance persist, irrespective of various difficulties and uncertainties left in the input quantity -- the left part of the scattering amplitude.

This paper will provide further evidences on the existence of $N^*(890)$, by directly finding a pole in the $S$ matrix
element calculated from  the  $N/D$ method. Early studies on low energy $\pi N$ scatterings via $N/D$ method may be found in Ref.~\cite{Gasparyan:2010xz} and references therein. Nevertheless, no report on the possible existence of a subthreshold resonance is known in the literature of $N/D$ studies, to the best of our knowledge. In our practice of $N/D$ calculations no spurious poles on first Riemann sheet are found to emerge. Also an $N/D$ calculation  faithfully reproduces all input dynamical as well as kinematical branch point singularities. We therefore think the $N/D$ method is rather reliable. {However, the calculations in $N/D$ studies do generate spurious branch cuts and spurious poles on the second sheet, due to the truncation of numerical integrations. Nevertheless, their effects can be evaluated  to see that the sum of hazardous contributions be negligible in many cases.}

This paper is organized as  follows: section \ref{intro} is the introduction. In section~\ref{prelude} a brief introduction to the $N/D$ method is given with a solvable toy model calculation. Also in section~\ref{prelude} we afford a review on the production representation which is found very illuminating in understanding the complicated $N/D$ calculations. {A subtlety occurs when using the production representation in dealing with $\pi$N scatterings: the dispersion representation for the background contribution to the phase shift has to be modified, an additional contribution emerges which is however cancelled almost exactly by contributions from two virtual poles located near the end points of the cut caused by $u$ channel nucleon pole exchanges. We put all the related discussions with respect to the subtlety in the appendix.}
Section~\ref{s=0} focuses on the singularity structure of partial wave amplitudes at $s=0$, including the discussion on why chiral expansions break down here, and  on how high energy contributions enter through an analysis on regge asymptotic behavior of $T(s)$ when $s\to 0$. Section~\ref{N/D} devotes to numerical analyses on how can a subthreshold resonance emerge under various phenomenological inputs.
\section{The $N/D$ recipe, a prelude}\label{prelude}
\subsection{A brief introduction to $N/D$ method}
\begin{figure}[h]
\begin{center}
    \includegraphics[width=11cm]{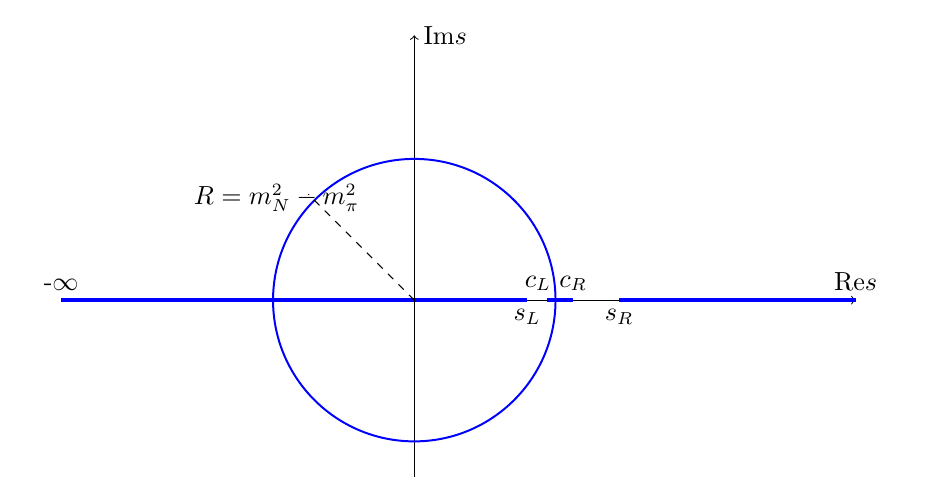}
    \caption{Branch cuts (thick blue lines) of partial wave $\pi$N elastic scattering amplitudes, {where $c_L = {(m_N^2-m_\pi^2)^2}/{m_N^2}$, $c_R = m_N^2+2m_\pi^2$.} }\label{piNcut}
    \end{center}
\end{figure}

The partial wave $T$ matrix element is expressed as
\be\label{1}
T={N}/{D}\ ,
\ee
where $D$ contains only the $s$-channel unitarity cut or the right hand cut $R$, whereas $N$ only contains the left hand cut ($l.h.c.$) or $L$ . See Fig.~\ref{piNcut}, $R=[s_R, +\infty)$ whereas $L$ represents all branch cuts except $R$ in Fig.~\ref{piNcut}. In section~\ref{221} we will briefly review on how to determine the cut structure in Fig.~\ref{piNcut}~\cite{Kennedy:1961}.

Partial wave unitarity in  single channel approximation
\bqa\label{PU}
\mathrm{Im}_R T(s)=\rho(s)|T(s)|^2\ ,
\eqa
{where
\be\label{rhos}\rho=\sqrt{(s-s_L)(s-s_R)}/s\ee
 with  $s_L=(m_N-m_\pi)^2$, $s_R=(m_N+m_\pi)^2$,}
leads to the following relations
\bqa\label{Dim}
\mathrm{Im}_R[D(s)] &=& -\rho(s)N(s)\ ,\no
\mathrm{Im}_L[N(s)] &=& D(s) \mathrm{Im}_L[T(s)]\ ,
\eqa
and subsequential  $N/D$ equations:
\be\label{DR}
\begin{split}
& D(s) = 1 - \dfrac{s-s_0}{\pi} \int_{R}\dfrac{\rho(s^\prime)N(s^\prime)}
{(s^\prime-s)(s'-s_0)}ds^\prime \ , \\
&N(s) =N(s_0) +\dfrac{s-s_0}{\pi}\int_L\dfrac{ D(s^\prime) \mathrm{Im}_L[T(s^\prime)]}
{(s^\prime-s)(s^\prime-s_0)}ds^\prime\ .
\end{split}
\ee
{Notice that when there appears circular cut in $T$ on $s$ plane as shown in Fig.~\ref{piNcut}, the second equation of Eq.~(\ref{Dim}) should be read as $\mathrm{disc}[N(s)] = D(s) \mathrm{disc}[T(s)]$. Also in the second equation of Eq.~(\ref{DR}), when the integration is performed on the circular cut which belongs to a subset of $L$, $\mathrm{Im}_L[T(s)]$ should be replaced by $\frac{1}{2i}\mathrm{disc}[T(s)]$.}
{After getting a numerical solution of Eq.~(\ref{DR}), analytical continuation to the complex plane is straightforward: taking $s$ to be complex while evaluating the integration in Eq.~(\ref{DR}) when $s$ is in the first sheet, and taking
 \be
 D^{\mathrm{II}}(s)=D(s)+2i\rho N(s)\ ,\,\,\, N^{\mathrm{II}}(s)=N(s)\ ,
 \ee
 when $s$ lies on the second sheet.}
 In Eq.~(\ref{DR}), the left cut of the partial wave $T$ matrix element, $\mathrm{Im}_LT$ is an input quantity. Throughout this paper, we only discuss $\mathrm{Im}_LT$ extracted from tree level amplitudes. Hence, except in the case with $t$-channel $\rho$ exchange as discussed in section~\ref{rho}, where an arc cut in $s$ plane will be met, {$L$ is always on the real axis. For example, {for pure tree level chiral amplitudes, {$L=(-\infty, 0] \cup [c_L,c_R]$.}} We will make a rather detailed discussion on how to determine  $l.h.c.$s in section~\ref{221}.

 To solve the integral equations one may substitute  $D$ into the second equation of Eq.~(\ref{DR}) to get
\begin{equation}\label{Nint}
N(s) = N(s_0) +\tilde B(s,s_0) +\dfrac{s-s_0}{\pi}\int_R \dfrac{\tilde B(s^\prime,s)\rho(s^\prime)N(s^\prime)}{(s^\prime-s_0)(s^\prime-s)}ds^\prime\ ,
\end{equation}	
	with
\be\label{Nint'}
\tilde B(s^\prime,s) =\dfrac{s^\prime-s}{2\pi i}\int_L\dfrac{\mathrm{disc}T(\tilde s)}{(\tilde{s} -s)(\tilde{s}-s^\prime)}d\tilde{s}\ ,
\ee
and use the inverse matrix method to obtain a numerical solution. {Throughout this paper, we set $s_0=1$ GeV$^2$, a value a little bit below the elastic threshold $s_R$.

As we will see later in this paper, there exists a subtlety when using Eq.~(\ref{DR}) to discuss un-equal mass scatterings. The problem comes from a singularity at $s=0$ in the partial wave amplitude and in its left cut, which stems from
high energy region $t\to +\infty$ through partial wave projections, and from relativistic spin kinematics as well. But before dealing with this problem, we are more eager in finding a solution of Eq.~(\ref{DR}) in a simplified toy model, in the following subsection. 

\subsection{A toy model calculation}\label{toy}
In $N/D$ recipe the input quantity is $\mathrm{Im}_L\, T$. Let us begin with a simple version of $N/D$ study by assuming
$\mathrm{Im}_L\, T$ being simulated by a set of Dirac $\delta$ functions, {or equivalently
\be\label{gamma}
\, N=\sum_i\frac{ \gamma_i}{s-s_i}\ ,
\ee
which is to be used in the first equation of Eq.~(\ref{DR}).
I.e., there is no need for a subtraction in the second equation of Eq.~(\ref{DR}).}
The $T$ matrix in such a situation is analytically solvable. {We (arbitrarily) choose $Case$ I: one pole at $s_1=0$, and $Case$ II: one pole at $s_1=-m_N^2$, }and fit to the ``data" obtained from the solutions of Roy Steiner equations~\cite{Hoferichter_2016} by tuning the parameter $\gamma_1$, and search for poles on the $s$-plane. Both cases give a good fit to the data, and a sub-threshold pole emerges in each case with a  location  listed in Table~\ref{poleposition}.
\begin{table}[h]
    \centering
    \begin{tabular}[width=12cm]{|c|c|c|}
        \hline
          & $Case$ I & $Case$ II \\
            \hline
        $s_1$ &  0 & $-m_N^2$ \\
          \hline
        $\gamma_1$ (GeV$^2$) &  0.79 & 1.34 \\

    \hline
        $\sqrt{s_{pole}}$(GeV) &  0.810 - 0.125i & 0.788 - 0.185i \\      \hline
    \end{tabular}
    \caption{Subthreshold pole locations using input Eq.~(\ref{gamma}).}\label{poleposition}.
\end{table}
The phase shift and its PKU decomposition~\cite{Zheng:2003rw}\cite{Zhou:2006wm} are plotted in Fig.~\ref{polepkuphase}.  On the left of Fig.~\ref{polepkuphase}, we see the familiar picture that the background cut contribution to the phase shift is concave and negative while the subthreshold resonance pole provides a positive and convex phase shift above threshold to counterbalance the former contribution, and the sum of the two reproduces the steadily rising phase shift data. In order to have a better  understanding of this phenomenon, a brief introduction to the production representation of partial wave elastic scattering $S$ matrix element is needed.
\begin{figure}[h]
\begin{center}
    \includegraphics[width=7cm]{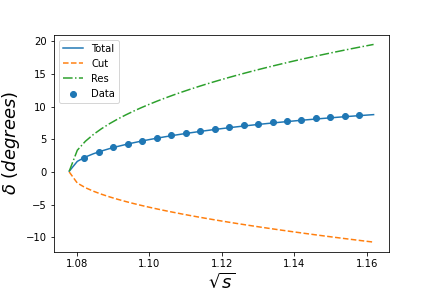}
    \includegraphics[width=7cm]{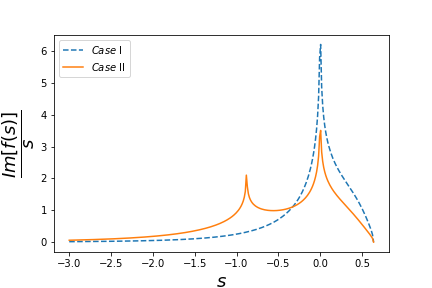}
    \caption{left: fit to the $S_{11}$ channel $\pi$N scattering phase shift data, taking $Case$ II as an example (data from Ref.~\cite{Hoferichter_2016}); right: the spectral function $\mathrm{Im}_L f(s)/s$ of $Case$ I and II. Notice that the singularity at $s=0$ in $Case$ II is due to the kinematical singularity in $\rho(s)$ defined {in Eq.~(\ref{rhos}) } rather than dynamical. For the definition of different contributions to the phase shift, one is referred to Sec.~\ref{PKU}.}\label{polepkuphase}
    \end{center}
\end{figure}

\subsection{The PKU representation}\label{PKU}
 The ``spectral" function drawn on the $r.h.s.$ of Fig.~\ref{polepkuphase} is defined as following:
\bqa\label{ffun}
&&f(s) = \dfrac{s}{{2i}\pi}\int_L ds^\prime\dfrac{\mathrm{disc}_Lf(s^\prime)/s^\prime }{(s^\prime-s)}+\dfrac{s}{{2i}\pi}\int_R ds^\prime\dfrac{\mathrm{disc}_Rf(s^\prime)/s^\prime }{(s^\prime-s)}\ ,\\
&&\mathrm{disc}_{L,R}f(s) = \mathrm{disc}_{L,R}\left(\dfrac{\ln S(s)}{2i\rho(s)}\right)\ ,
\eqa
where the partial wave $S$ matrix element $S=1+2i\rho T$ and $\rho$ is the kinematic factor. {In Eq.~(\ref{ffun}), the integration domain is depicted in Fig.~\ref{piNcut}, it is noticed that $R=[(m_N+m_\pi)^2,+\infty)$ but actually the integrand vanishes in the elastic region, or in other words, $R$ starts from first inelastic threshold.}
Elastic partial wave $S$ matrix elements satisfy a production representation like follows:
\be\label{ffun'}
S=\prod_iS_i\times e^{2i\rho(s)f(s)}\ .
\ee
{Detailed discussions on how to obtain Eq.~(\ref{ffun'}) can be found in Refs.~\cite{Zheng:2003rw}\cite{Zhou:2006wm}.}
 The production representation exhibits some nice features which are very useful in our analyses. One advantage is that   phase shifts from different sources are additive, {i.e.,
 \be\label{delta}
 \delta(s)=\sum_i\delta_i+\delta_{cut}\ ,
 \ee
  where $\delta_i$ comes from the $i$-th pole contribution described by $S_i$ in Eq.~(\ref{ffun'}). For a resonance located at $z_0$ ($z_0^*$) it is
  \be
  \delta_{res}=\arctan\[\frac{\rho(s)sG[z_0]}{M^2(z_0)-s}\]\,\,\,(S_R=e^{2i\delta_R})\ ,
  \ee
  with $G(z_0)=\mathrm{Im}[z_0]/\mathrm{Re}[z_0\rho(z_0)]$ and $M^2(z_0)=\mathrm{Re}[z_0]+\mathrm{Im}[z_0]\mathrm{Im}[z_0\rho(z_0)]/\mathrm{Re}[z_0\rho(z_0)]$. For a virtual state located at $s_0$ ($s_L<s_0<s_R$), its contribution is
  \be
  \delta_v=\arctan\[{\frac{\rho(s)s}{s-s_L}\sqrt{\frac{s_0-s_L}{s_Rs_0}}}\]\,\,\,(S_v=e^{2i\delta_v})\ .
  \ee
  For a bound state, the above expression changes sign, i.e., $\delta_b=-\delta_v$ ($S_b=e^{2i\delta_b}$).} { The last contribution in Eq.~(\ref{delta}), i.e., the background contribution to the phase shift, is
  \be
  \delta_{cut}=\rho(s)f(s)\ ,
  \ee
  with function $f(s)$ defined in Eq.~(\ref{ffun}).}

   { The above description of the production representation only repeats what have been established in the literature. In our study of $\pi$N scatterings, however, an unexpected new phenomenon occurs. Simply speaking,
 the cut structure of the background phase shift, $\delta_{b.g.}$, has to be modified, due to the presence of the $u$--channel nucleon exchange.~\footnote{The cut structure of $S_{b.g.}=\exp{2i\rho f(s)}$ remains unchanged, as it should.} As a result, an additional contribution exists in the expression of $\delta_{b.g.}$, which is, however, almost exactly cancelled by the contributions from two newly found virtual poles located near the end points (i.e., $c_L$ and $c_R$ defined in Fig.~\ref{piNcut}) of the segment cut induced by the $u$ channel nucleon exchange. Since the explanation is rather lengthy, we postpone the discussions in the appendix.}

  The  property that the phase shift contributions are additive is vital in tracing the physical origin of the phase shift.
As having been stressed repeatedly, the positive value of  $\mathrm{Im}_L f(s)/s$ guarantees that the background contribution to the phase shift be negative and concave, and hence an isolated singularity on the second sheet is needed to take charge of the steady rise of near threshold phase shift. {The {circular cut} caused by $t$-channel exchanges and the $u$-channel cut ({combined with effects of two virtual poles}) are not considered yet here, nevertheless they are numerically small  as comparing with the cut lying on $(-\infty, s_L]$.}\footnote{Previous examples can be found in Refs.~\cite{Zheng:2003rw}\cite{Wang:2018nwi}\cite{Ma:2020sym}.}
Hence a large and positive $\mathrm{Im}_L f(s)/s$ at $s=0$, looks  important as suggested in Fig.~\ref{polepkuphase}. The strong enhancement of $\mathrm{Im}_L f(s)/s$ at $s=0$ is due to two reasons: one is from the kinematic singularity at $s=0$ from Eq.~(\ref{rhos}), the other  is from the possible singularity in $T(s)$ when $s\to 0$.  An example for the former is fit $Case$ II in section~\ref{toy} where $T(0)\sim \mathrm{constant}$, while the latter example is provided by fit $Case$ I  where $T(s)\sim O(s^{-1})$ when $s\to 0$. In general, making use of the property of real analyticity for $S$ matrix elements, one recasts $\mathrm{Im}_L f(s)$, {when $s$ lies on the real axis,} as
$-\ln{|S(s)|}/2\rho(s)=-\ln{[1-4\rho(s)\mathrm{Im}_LT(s)+4\rho(s)^2|T|^2]}/4\rho(s)$. Hence if $T(s)$ does not vanish when  $s\to 0$, then $\ln{|S(s)|}$ diverges logarithmically, since $\rho(s)\propto s^{-1}$  at origin. It may be worth stressing that the singularities caused by relativistic kinematics truly exist and bring physical consequences, since they enter  the physical equations such as Eq.~(\ref{PU}). A good example to realize this comes from {figure~5 and Eq.~(55)} of Ref.~\cite{Xiao:2000kx}: without the kinematical singularity the data curve can simply not be explained.

On the other side, the inelastic right hand cut  contribution to the phase shift in Eq.~(\ref{ffun}) should always be positive~\cite{Wang:2017agd}, since in that region $|S(s)|=\eta<1$, with $\eta$ the inelasticity parameter. In $N/D$ calculations, a cutoff to the integral interval has to be adopted, i.e., [$s_R, +\infty)$ to be replaced by $[s_R,\Lambda_R^2]$. In this situation, it is not difficult to understand that the truncated $N/D$ integration actually {violates unitarity by introducing  a spurious branch cut starting from $s=\Lambda_R^2$,} in the sense that the effective $\eta$ parameter exceeds unity: ${|S(s)|^2}={[1-4\rho(s)\mathrm{Im}_RT(s)+4\rho(s)^2|T|^2]}=[1+4\rho(s)^2|T|^2]>1$, when $s>\Lambda_R^2$, since a truncation implies actually $\mathrm{Im}_RT=0$.
So one has to be cautious when performing the $N/D$ calculations by monitoring to what extent unitarity is violated. This may be fulfilled at quantitative level by, for example, {calculating the contribution from the region above $\Lambda^2_R$ to the phase shift, through Eq.~(\ref{ffun}). {It is found that, when doing calculations in this paper, the violation can either be large or  small, depending on whether or not the input quantity $T_L(s)$ at $s=\Lambda_R^2$ is too large or small. For the former, an example is the $\chi$PT input which is not valid anymore at $\Lambda_R^2=1.48$ GeV$^2$, i.e., the value we choose {in most of} our calculations. However, the $N^*(890)$ exists with a rather stable location, irrespective to the pollution of truncation of integration.}

Encouraged by the discussions made in  section~\ref{toy}, we plan to make a more realistic $N/D$ calculation at next steps. The first thing needs to be settled down is of course choosing an input $\mathrm{Im}_L T$ as much realistic as possible. One may choose the $\chi$PT outputs as an input here, as what is adopted in Refs.~\cite{Wang:2018nwi}\cite{Wang:2018gul}\cite{Wang:2017agd}. A careful analysis reveals, however, that the partial wave projection of $\chi$PT amplitudes encounters  a rather severe problem at $s=0$.
In the following section we dedicate to the study of such a problem.

\section{The singularity structure of $T(s)$ at $s=0$}\label{s=0}
As suggested in the end of the above section, to make a more realistic calculation, one may use $\chi$PT amplitudes to extract  $\mathrm{Im}_L T$~\cite{Wang:2017agd}.
However, the  results may not be directly applicable to Eq.~(\ref{DR}), and should be treated with great care.  The integration interval for $N(s)$ is $L=(-\infty,0]$ {for tree level estimation here}.  The $\mathcal{O}(p^3)$ level $\mathrm{Im}_L T$  behaves as $\mathcal{O}(s^{-7/2})$ when $s\to 0$, and at $\mathcal{O}(p^2)$ level it behaves as $\mathcal{O}(s^{-5/2})$, while from a rather general argument to be discussed in section~\ref{221} these strong singularities at $s=0$ are not physical. Hence one has to find a way to get rid of these artifacts caused by  partial wave projections of $\chi$PT amplitudes.

 In general, singularities of the partial wave $T$ matrix element at $s=0$ come from two aspects:
 \begin{itemize}
 \item the $\mathcal{O}(p^n)$ ($n\geq 2$) level $\chi$PT amplitudes behave as $s^{-n-1/2}$ when $s\to 0$ for unequal mass scatterings, after partial wave projections, and the integer $n$ increases when the chiral order increases.

	\item the left-hand cut around $s=0$ receives singular contribution from  high energy region of crossed channels, i.e., $t,u\to\infty$, through partial wave projections;
	
\end{itemize}
We will address these problems in some details below. It is worth pointing out beforehand, that though in $N/D$ approach one has to deal with these problems cautiously,  {the calculations made in Refs.~\cite{Wang:2017agd} and \cite{Wang:2018nwi} are luckily insensitive to the problem. Though the background contribution to the phase shift will be enhanced incorrectly by the contribution in the vicinity of $s=0$, it will be largely compensated by tuning the cutoff parameter $\Lambda_L^2$ when evaluating the $l.h.c.$ integral.} More interestingly even though the integral of $f(s)$ is enhanced near $s=0$, its derivatives behave very differently. {For example,
\be
\frac{d f(s)}{ds}=-\dfrac{1}{\pi}\int ds^\prime\dfrac{\ln{|S(s')|}/2\rho(s')}{(s^\prime-s)^2}\ .
\ee
Now the integrand near $s^\prime=0$ behaves like $\sim s^\prime\ln s^\prime$ and the unwanted singularity is killed. This fact may be translated into a more transparent language: in Refs.~~\cite{Wang:2017agd}\cite{Wang:2018nwi}, except the scattering length which anyway needs to be fitted by tuning $\Lambda^2_L$, the other quantities such as the curvature of the phase shift curve are quite immune to the disease infected from $s=0$.}
\subsection{Artificial singularities in partial wave $\chi$PT amplitudes}\label{section3.1}

We notate the process as $N(p,\sigma)+\pi(q)\to N(p',\sigma')+\pi(q')$, where $p,\ q, \ p',\ q'$ are the momenta and $\sigma,\ \sigma'$ are the spins. The Mandelstam variables are
\begin{equation}
	s=(p+q)^2=(p'+q')^2\ ,\quad t=(p-p')^2=(q-q')^2\ ,\quad u=(p-q')^2=(p'-q)^2\ .
\end{equation}
The full amplitude, ${\cal T}$, can be decomposed as (the following discussions are all for isospin $I=1/2$ only),
\begin{equation}\label{13}
	{\cal T}=\bar{u}(p',\sigma')\left[A(s,t)+\frac{\slashed{q}+\slashed{q}'}{2}B(s,t)\right]u(p,\sigma)\ .
\end{equation}
 The results of the scalar functions $A,B$ are for example listed in  {Refs.~\cite{Wang:2017agd,Wang:2018nwi} (further references are found in Refs.~\cite{Alarcon:2012kn}\cite{Chen:2012nx}{\cite{Bruns:2010sv}}\cite{Siemens:2016hdi}\cite{Siemens:2016jwj}\cite{Siemens:2017opr}).} From $\mathcal{O}(p^2)$ on, the $\chi$PT lagrangian contains $4$-point $\pi\pi NN$ contact terms, which contributes to the scalar functions as ($C$ refers to constants),
\begin{equation}\label{op23full}
	A\big[\mathcal{O}(p^2)\big]\supset C(s-u)^2\ ,\quad A\big[\mathcal{O}(p^3)\big]\supset C(s-u)^3\ .
\end{equation}
More explicitly,
\begin{itemize}
\item{at $\mathcal{O}(p^1)$ (Born and contact diagrams):}
\begin{equation}\label{p1ABterm}
	\begin{split}
		&A_{1}=\frac{g^2m_N}{F^2}\ \mbox{, }\\
		&B_{1}=\frac{1-g^2}{F^2}-\frac{3m_N^2g^2}{F^2(s-m_N^2)}-\frac{m_N^2g^2}{F^2}\frac{1}{u-m_N^2}\ \mbox{; }
	\end{split}
\end{equation}
\item{at $\mathcal{O}(p^2)$ (only contact diagram):}
\begin{equation}\label{p2ABterm}
	\begin{split}
		&A_2=-\frac{4c_1 m_\pi^2}{F^2}+\frac{c_2(s-u)^2}{8m_N^2F^2}+\frac{c_3}{F^2}(2m_\pi^2-t)-\frac{c_4(s-u)}{F^2}\ \mbox{, }\\
		&B_2=\frac{4m_Nc_4}{F^2}\ \mbox{; }
	\end{split}
\end{equation}
\item{at $\mathcal{O}(p^3)$ (Born diagram):}
\begin{equation}\label{p3ABtermB}
	\begin{split}
		&A_{3B}=-\frac{m_Ng}{F^2}\times 4m_\pi^2(d_{18}-2d_{16})\ \mbox{, }\\
		&B_{3B}=\frac{4m_\pi^2g(d_{18}-2d_{16})}{F^2}\times\frac{su+m_N^2(2u-3m_N^2)}{(s-m_N^2)(u-m_N^2)}\ \mbox{; }
	\end{split}
\end{equation}
\item{at $\mathcal{O}(p^3)$ (contact diagram):}
\begin{equation}\label{p3ABtermC}
	\begin{split}
		&A_{3C}=-\frac{(d_{14}-d_{15})(s-u)^2}{4m_NF^2}+\frac{(d_1+d_2)}{m_NF^2}(s-u)(2m_\pi^2-t)\\
&\,\,\,\,\,\,\,\,\,\,\,+\frac{d_3}{8m_N^3F^2}(s-u)^3+\frac{4m_\pi^2 d_5}{m_NF^2}(s-u)\ \mbox{, }\\
		&B_{3C}=\frac{(d_{14}-d_{15})(s-u)}{F^2}\ \mbox{. }
	\end{split}
\end{equation}
\end{itemize}
In the expressions above, $g$ is the axial-vector coupling constant, $F$ is the pion decay constant, and {$c_i$, $d_i$ are low-energy constants.}

The partial wave projection is done  on helicity amplitudes:
\begin{equation}\label{helamp}
	\begin{split}
		&{\cal T}_{++}=\sqrt{\frac{1+z_s}{2}}\big[2m_NA(s,t)+(s-m_\pi^2-m_N^2)B(s,t)\big]\ \mbox{, }\\
		&{\cal T}_{+-}=\sqrt{\frac{1-z_s}{2s}}\big[(s-m_\pi^2+m_N^2)A(s,t)+m_N(s+m_\pi^2-m_N^2)B(s,t)\big]\ \mbox{, }
	\end{split}
\end{equation}
where $z_s=\cos\theta$, and $\theta$ is the scattering angle; $T_{++}$ stands for that the helicities of the initial and final nucleon are both $+1/2$, while $T_{+-}$ means the final nucleon has helicity $-1/2$ . The relations between the Mandelstam variables and the scattering angle ($z_s=\cos\theta$) are:
\begin{align}
	&t(s,z_s)=2m_\pi^2-\frac{(s+m_\pi^2-m_N^2)^2}{2s}+\frac{[s-(m_\pi+m_N)^2][s-(m_\pi-m_N)^2]}{2s}z_s\ ,\label{tsz}\\
	&u(s,z_s)=m_\pi^2+m_N^2-\frac{s^2-(m_\pi^2-m_N^2)^2}{2s}-\frac{[s-(m_\pi+m_N)^2][s-(m_\pi-m_N)^2]}{2s}z_s\label{usz}\ .
\end{align}
The $S_{11}$ amplitude is from $J=1/2$ partial wave helicity amplitudes:
\begin{equation}\label{TppTpmpw}
	\begin{split}
	&T_{++}^J=\frac{1}{32\pi}\int_{-1}^1 dz_s {\cal T}_{++}(s,t(s,z_s)) d^J_{1/2,1/2}(z_s)\ \mbox{, }\\
	&T_{+-}^J=\frac{1}{32\pi}\int_{-1}^1 dz_s {\cal T}_{+-}(s,t(s,z_s)) d^J_{-1/2,1/2}(z_s)\ \mbox{, }
	\end{split}
\end{equation}
where $d$ stands for Wigner small-$d$ matrix. For $S_{11}$ channel,
\begin{equation}
T(S_{11})=T_{++}^{J=1/2}+T_{+-}^{J=1/2}\ .
\end{equation}
From this formula the singularity at $s=0$ is obvious. On the one hand, in Eq.~\eqref{helamp} the kinematic effects give an $s^{-1/2}$ factor, which makes $s=0$ a branch point. On the other hand, see for example in Eq.~~\eqref{p3ABtermC}, contact terms from $\chi$PT expansions lead to higher and higher order polynomials of $s-u$:
\begin{equation}
	{\cal T}\big[\mathcal{O}(p^n)\big]\supset C(s-u)^n\ .
\end{equation}
According to Eq.~\eqref{usz}, $u(s\to0)\to s^{-1}$, so {when $n\geq 2$}
\begin{equation}\label{op23}
	{\cal T}\big[\mathcal{O}(p^n)\big](s\to0)\sim C s^{-n-1/2} \ ,
\end{equation}
{where the extra factor $-1/2$ in the power comes from kinematic effects of helicity basis. }
Eq.~(\ref{op23}) indicates that the higher order of $\chi$PT calculation is employed, the stronger singularity will occur near $s=0$ and the chiral series breaks down.  Apparently this is only an artificial problem caused by  chiral expansions since it contradicts the genuine singularity structure expected when $s\to 0$, as discussed in the end of section~\ref{221}: {the Froissart bound forbids a power behavior like  Eq.~(\ref{op23})  when $s\to 0$.
In principle one would not hope Eq.~(\ref{op23}) to appear in the expression of $\mathrm{Im}_L T$ when using $N/D$. }

\subsection{High energy contributions from crossed channels}
\label{221}

One writes a spectral representation of the partial wave amplitude \cite{Kennedy:1961}, for $t$-channel:
\begin{equation}
	T\sim \int_{\sigma_t}^{+\infty}dt' \Sigma(s,t')\int_{-1}^1 d\cos\theta \frac{R(\cos\theta)}{t-t'}\ ,
\end{equation}
where $\Sigma$ is the {Mandelstam spectral function}, $\sigma_t=4m_\pi^2$ is the threshold of $t$-channel process, and $R$ is the basis function of the partial wave projection (usually the linear combinations of Wigner-$d$ matrices). The function $R$ can be expanded at $t=t'$: $R=R_0+(t-t')R_1+\cdots$, and only the leading order causes singularities:
\begin{equation}
	T\propto\int_{\sigma_t}^{+\infty}dt' \Sigma(s,t')\beta^{-1}\ln\left[\frac{\alpha+\beta}{\alpha-\beta}\right]\ ,
\end{equation}
with
\begin{equation}
	\alpha=t'-2m_\pi^2+\frac{(s+m_\pi^2-m_N^2)^2}{2s}\ ,\quad \beta=\frac{[s-(m_\pi+m_N)^2][s-(m_\pi-m_N)^2]}{2s}\ .
\end{equation}
Therefore the left-hand cut is described by the equation
\begin{equation}
	\alpha^2=\beta^2\ ;
\end{equation}
which gives
\begin{equation}\label{solt}
	s_\pm(t')=m_\pi^2+m_N^2-\frac{t'}{2}\pm\frac{1}{2}\sqrt{(t'-4m_\pi^2)(t'-4m_N^2)}\ .
\end{equation}
When $t'$ takes the value from $\sigma_t$ to $+\infty$, the trajectory of this solution traces out the left-hand cut. The following conclusions can be obtained:
\begin{itemize}
	\item when $t'\in [4m_\pi^2,4m_N^2]$, the cut appears as a circle $\text{Re}s^2+\text{Im}s^2=(m_N^2-m_\pi^2)^2$, and the endpoint to the right {$s=m_N^2-m_\pi^2$} corresponds to $t'=\sigma_t=4m_\pi^2$;
	\item when $t'\in (4m_N^2,+\infty)$, $s_-$ generates the cut $(-\infty,m_\pi^2-m_N^2)$, and $-\infty$ corresponds to $t'\to+\infty$;
	\item {when $t'\in (4m_N^2,+\infty)$, $s_+$ generates the cut $(m_\pi^2-m_N^2,0)$, and actually $0$ corresponds to $t'\to+\infty$;}
\end{itemize}
One performs a similar analysis in $u$-channel and the solution is
\begin{equation}\label{solu}
	s_1(u')=\frac{(m_\pi^2-m_N^2)^2}{u'}\ ,\ s_2(u')=2(m_\pi^2+m_N^2)-u'\ .
\end{equation}
There is a nucleon pole $u'=m_N^2$, giving a segment cut $((m_\pi^2-m_N^2)^2/m_N^2, 2m_\pi^2+m_N^2)$. When $u'>\sigma_u=(m_\pi+m_N)^2$, $s_1$ gives $(0,(m_N-m_\pi)^2)$ and $s_1\to0$ just when $u'\to+\infty$; while $s_2$ generates $(-\infty,(m_N-m_\pi)^2)$ with $s_2\to-\infty$ when $u'\to+\infty$.

It was pointed out in Ref.~\cite{Zhou:2006wm} that, for meson -- meson scatterings, if $T(s,t)\sim O(t^n)$ when $t\to \infty$ for fixed $s$, then the partial wave amplitude behaves as $T(s)\sim O(s^{-n})$ when $s\to 0$.   Considering the Froissart bound in the crossed channel $|T(t,\cos\theta_s=1)|< t\ln^2t$, it is expected that the proper singularity behavior for $s$-channel partial wave amplitude near $s=0$ is no more singular than {$T(s)\sim O(s^{-1})$ }(up to some logarithmic corrections).
 This estimation can even further be improved.
It is seen from the above discussions that as $s\to 0_-$, there is a high energy contribution from the $t\to +\infty$ region. In this region the full amplitude is governed by $t$-channel ($\pi\pi\to N\bar N$) reggeon exchanges. The leading Regge trajectory is  $\Delta(1232)$ with the intercept parameter $\alpha_\Delta(0)\simeq 0.19$~\cite{Jakob:1969hn}, which leads to a weak singularity
\be T\sim s^{-\alpha_\Delta(0)}\ ,\ee
for the partial wave amplitude, when $s\to 0_-$. The $s\to 0_+$ limit is the same {according to for example Ref.~\cite{Jakob:1969hn}.}

{ All the discussions given above in this section are on how to determine $l.h.c.$s generated dynamically, or in other words, cuts originated from physical absorptive singularities from crossed channels. Besides these dynamical $l.h.c.$s there exists an additional cut $(-\infty, 0]$ for pure kinematical reasons: the nucleon spinor wave function provides a $\sqrt{s}$ branch cut which already showed up in the second equality in Eq.~(\ref{TppTpmpw}). The effect of branch cut singularity from  relativistic kinematics truly exists, as has already been addressed in section~\ref{PKU}.}

\section{Numerical analyses of $N/D$ method}\label{N/D}
It has been made clear in the above section, that using $\chi$PT inputs of $\mathrm{Im}_LT$ encounters the problem that the partial wave projections of  $\chi$PT amplitudes lead to a strong but incorrect singularity at $s=0$, by violating what is  expected from rather general constraints of quantum field theory.  Nevertheless it is not clear yet, to what extent the use of $\chi$PT results may distort the physical output. In the following we devote to the study of this problem by invoking $\mathcal{O}(p^2)$ and $\mathcal{O}(p^3)$ (tree level amplitude only) $\chi$PT results, since in $\mathcal{O}(p^1)$ case no free parameter is available for a data fit. Nevertheless, {the $\mathcal{O}(p^1)$ $N/D$ unitarization} can still be made and compared with data, which ends up with a pole location $\sqrt{s}=1.08-i0.23$ GeV and a steadily rise phase shift larger than data by roughly 5 degrees at $\sqrt{s}=1.16$ GeV.

\subsection{$N/D$ calculations using pure $\chi$PT inputs}
{
The singularity of $\mathcal{O}(p^2)$ $\mathrm{Im}_LT$ when $s\to 0$ behaves as $\mathcal{O}(s^{-2-1/2})$. Which, as discussed previously, is not physical. We nevertheless still perform the $N/D$ calculation to see what happens. In doing such a calculation it is noticed that the sick singularity behavior destroys the validity of Eq.~(\ref{Nint'}). To overcome the problem,  the auxiliary function $\tilde B$ in Eqs.~(\ref{Nint}) and (\ref{Nint'})
can be formally written as $\tilde B(s,s_0)=T_L(s)-T_L(s_0)$, where $T_L$ is taken as the $O(p^1)$ partial wave amplitude (Eq.~(\ref{op1})) plus the $O(p^2)$ part of $T^J_{+-}$, since at $O(p^2)$ level $T^J_{++}$ does not contribute to the imaginary part on the left.
In this way we avoid the discussions on possible subtractions encountered at  two endpoints of the integral defined in Eq.~(\ref{Nint'}).\footnote{One may redefine $T(s)=\bar T(s)/s^2$ and $\bar T(s)={N(s)}/{D(s)}$ to avoid the singularity in the integral in Eq.~(\ref{Nint'}) at $s=0$. The results are similar to  analyses presented in this manuscript.}
The subtraction constant $N(s_0)$ appeared in Eq.~(\ref{Nint}), {or $T(s_0)$,} serves as a free fit parameter.}

 At $\mathcal{O}(p^2)$ level there are four low energy constants (LECs) $c_i$ with $i=1,\cdots,4$.
There are different sets of $c_i$ parameters found in the literature ({e.g., Refs.~\cite{Hoferichter:2015tha}},\cite{Gasparyan:2010xz},\cite{Yao:2016vbz},\cite{Scherer:2012xha},\cite{Wang:2018nwi}).
For these LECs, certain  bounds, i.e., the positivity constraints~\cite{Sanz-Cillero:2013ipa} {are obeyed.}

A good fit is obtained with 
{$c_1=-0.40,\ c_2=3.50,\ c_3=-3.90,\ c_4=2.17$,  $N(s_0)=0.47$,} and the pole locates at
 \be
\label{op2pole}\sqrt{s}=1.01\pm 0.19\ \mathrm{GeV}.
\ee
{In addition, we have also employed different sets of $c_i$, and the results change very little. {For instance when we take the central values of $c_i$'s from Ref.~\cite{Hoferichter:2015tha} (NLO): $c_1=-0.74\pm0.02,\ c_2=1.81\pm0.03,\ c_3=-3.61\pm0.05,\ c_4=2.17\pm0.03$ (in units of GeV$^{-1}$), the pole position is $0.99-0.16i$ GeV though the phase shift does not fit data very well. } {It is noticed that,  the spurious branch cut becomes a bit annoying here, contributing to the phase shift at $\sqrt{s}=1.16$ GeV roughly {-6$^\circ$}. Nevertheless, this spurious effect is far from being dominant as can be seen from Fig.~\ref{ndphase}, where one finds the total background contribution exceeds -30$^\circ$, and hence is not worrisome.}} It is not totally clear to us yet  why and under what situation the spurious branch cut  becomes numerically visible. One conjecture is that the $\mathcal{O}(p^2)$ $\chi$PT input itself becomes {sick at $\Lambda_R^2=1.48$ GeV$^2$,} hence amplifies the contribution of the spurious branch cut . {Finally, since the effect generated from the cut at $s=\Lambda_R^2$ is small, we still think such type of solutions are acceptable  for the evaluation of physics at lower energies.}

 As a comparison we plot the ``spectral function", i.e., ${\mathrm{Im}_Lf(s)}/{s}$, obtained here together with those discussed in section~\ref{toy}, in Fig.~\ref{ndphase}.
\begin{figure}[h]
\begin{center}
    \includegraphics[width=7cm]{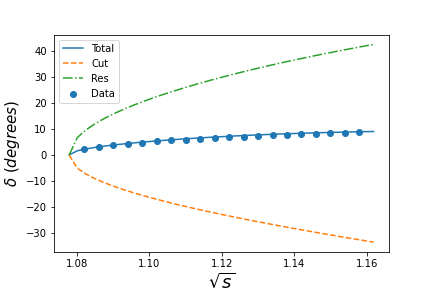}
    \includegraphics[width=7cm]{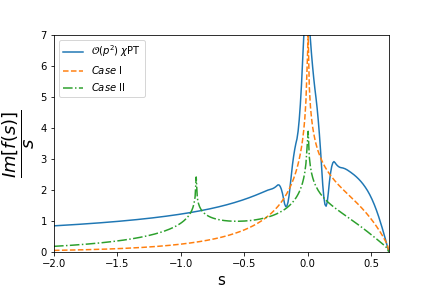}
    \caption{left: $N/D$ fit to the $S_{11}$ phase shift with $\mathcal{O}(p^2)$ $\chi$PT input; right: a comparison of different ${\mathrm{Im}f(s)}/{s}$ in different situations. }\label{ndphase}
    \end{center}
\end{figure}
Comparing Eq.~(\ref{op2pole}) with that in table~\ref{poleposition}, and different $l.h.c$ contributions drawn in Fig.~\ref{ndphase},
we observe that the $\mathcal{O}(p^2)$ calculation overestimates the $l.h.c$ contribution comparing with that of $Case$ II. As a result, the pole contribution to the phase shift has to be increased and hence the pole location  has to move towards to the right direction in  $s$ plane closer to the $\pi N$ threshold. But of course, such discussions are only meaningful under the condition that the effects of spurious branch cut  and the spurious pole around $s=\Lambda^2_R$ cancel each other.

If all the spurious contributions are ignored,\footnote{Because of two excuses: firstly they cancel each other; secondly, they are from distant places  as they are associated with the cutoff at $s=\Lambda^2_R$ anyway. It is desirable to know the possible origin of spurious effects since it leaves the hope to isolate and to remove them. On the contrary, it is difficult to cure the similar problem in Pad\'{e} approximation\cite{Qin:2002hk}.   } then we may ask a question. The calculation here and that in Ref.~\cite{Wang:2017agd} use the same `data' sample and $l.h.c.$ while in Ref.~\cite{Wang:2017agd} the pole locates at $\sqrt{s}=0.86\pm 0.05- i(0.13\pm0.08)$ GeV. Comparing with Eq.(\ref{op2pole}) it is found that there exists a rather large systematic error in determining the pole location. One possible reason to explain this may come from the fact that in Ref.~\cite{Wang:2017agd} a truncation of $l.h.c.$ is performed while in here there is no truncation on the left, {see Eq.~(\ref{Nint})}. In fact in the calculation made in table~4 of Ref.~\cite{Wang:2017agd}, it is found that when sending the cutoff $s_c$ to $\infty$ the pole location moves upwards to $\sqrt{s}=0.91- i0.21$ GeV, i.e., getting closer to Eq.~(\ref{op2pole}).

Finally, we have also tested the $\mathcal{O}(p^3)$ inputs (at tree level only) and found that the outputs are  similar to the situation found in the $\mathcal{O}(p^2)$ case, so we no longer discuss the results here anymore.

\subsection{$N/D$ calculation using phenomenological models}\label{rho}

 In the above section we have made discussions on the problem encountered  when using $\chi$PT results to estimate the $l.h.c$. The higher order terms appeared in the chiral lagrangian describing $\pi$N interactions are obtained by integrating out heavy degrees of freedom like the $\rho$ meson in  $t$-channel and $N^*$ resonances in  $s$ and $u$-channels, etc..~\footnote{In the meson--meson scattering lagrangian, the LECs at $\mathcal{O}(p^4)$ level are known to be saturated by heavy degrees of freedom~\cite{Ecker:1988te}. In meson--baryon system, systematic studies on this point are not known to the authors.} The ill singularities at $s=0$ in partial wave chiral amplitudes come at least partly from integrating out heavy degrees of freedom. To see this more clearly, let us write down an effective interaction lagrangian responsible for $t$-channel $\rho$ meson exchange,
\begin{align}
\mathcal{L}_{t} = g_{\rho}\vec{\rho}^\mu\cdot(\partial_\mu \vec{\pi}\times \vec{\pi})+g_{\rho}\bar{N}\dfrac{1}{2}\vec\tau\cdot(\gamma^\mu\vec{\rho}_\mu+\dfrac{\kappa}{2m_N}\sigma^{\mu\nu}\partial_\mu\vec{\rho}_\nu)N\ ,
\end{align}
where $g_{\rho}$ and $\kappa$ are resonance coupling constants, $\vec{\tau}$ are Pauli matrices, $\vec{\rho}_{\mu}$, $\vec{\pi}$ and $N$ refer to $\rho$ resonance, pion and nucleon, respectively.
For $S_{11}$ channel, the $\rho$ exchange contribution to the invariant amplitude, Eq.~(\ref{13}),  can be obtained:
\begin{equation}
A= \frac{g_{\rho}^2\kappa(u-s)}{2m_N(t-m_{\rho}^2)}\ ,\,\,\, B=\frac{2 g_{\rho}^{2}(\kappa+1)}{t-m_{\rho}^{2}}\ .
\end{equation}
Now, if a $1/m_{\rho}$ expansion is made, at   leading order we have
\begin{equation}
A=\frac{g_{\rho}^{2} \kappa(s-u)}{2 m_{N} m_{\rho}^{2}}\ ,\,\,\,
B=-\frac{2 g_{\rho}^{2}(\kappa+1)}{m_{\rho}^{2}}\ .
\end{equation}
Comparing with Eq.~(\ref{p2ABterm}), we find that  the $\rho$ meson exchange only contributes to $c_4$ term~\cite{Bernard:1996gq}. As we already know from the discussions made in section~\ref{section3.1}, the $c_4$ term will cause an $s^{-3/2}$ singularity after partial wave projection. This is avoidable, if we do not make a $1/m_{\rho}$ expansion in the beginning. It can be seen that all these resonance exchange amplitudes contain singularity of $s^{-1/2}$ type at most when $s=0$.
Therefore partial wave projections and chiral expansions do not commute, which can be checked directly by evaluating $t$-channel $\rho$ exchange contributions by making partial wave projections of Eqs.~(\ref{helamp}) and (\ref{TppTpmpw}).

{We further make an asymptotic expansion} of the $\rho$ exchange contributions to $T(S_{11})$ in the vicinity of $s=0$ and find that the first two most singular terms are of type
 \bqa\label{asymptotic}\frac{a+bs}{\sqrt{s}}\ ,
 \eqa
 which are not of type $\mathcal{O}(s^{-5/2})$ obtained if the $\rho$ propagator were expanded beforehand.
Similar things happen if we introduce, for example, a $u$-channel as well as $s$-channel $S_{11}$ resonance exchange.
In this situation one can prove that expanding the $N^*$ propagator in the full amplitudes leads  contributions to $c_3$ and $c_4$ terms in Eq.~(\ref{p2ABterm}). Not making a chiral expansion beforehand the resonance exchange contribution to the partial wave amplitude can be expanded at $s=0$ and similar results as Eq.~(\ref{asymptotic}) are again obtained, {so do the  $P_{11}$  resonance exchange contributions.}

Explicit expressions for  resonance contributions to parameters, e.g., $a$ and $b$ defined in Eq.~(\ref{asymptotic})  are obtainable. However, another obscure problem occurs here. These coefficients depend on the mass parameters of the exchanged resonances and do not vanish as the resonance mass gets large, {which seems to contradict the general expectation from the decoupling theorem~\cite{Appelquist:1974tg}\cite{Ovrut:1980eq}.} \footnote{The reason behind this phenomenon is that as $s\to 0$ the interval of partial wave integration diverges, make any ``heavy" mass scale not heavy anymore and cannot be naively integrated out.} Without a deeper understanding on this problem, we point out that {the sign of contributions from different sources can be different.} {For example, the $t$-channel $\rho$ exchange contribution to parameter $a$ is
 $ a(\rho)=-\frac{g_{\rho}^{2} \kappa\left(m_{N}^{2}-m_{\pi}^{2}\right)}{64 \pi m_{N}}$; the $\frac{1}{2}^{+}$ baryon exchange contribution is  $a(N^{*+})=\frac{\left(g_{N^{*}}\right)^{2}\left(-m_{N}^{2}+m_{\pi}^{2}\right)\left(3 m_{N}^{2}+\left(m_{N^{*}}\right)^{2}\right)}{128 \pi F^{2} m_{N^{*}}}$; whereas the $\frac{1}{2}^{-}$   resonance exchange contribution is $a(N^{*-})=\frac{\left(g_{N^{*}}\right)^{2}\left(m_{N}^{2}-m_{\pi}^{2}\right)\left(3 m_{N}^{2}+\left(m_{N^{*}}\right)^{2}\right)}{128 \pi F^{2} m_{N^{*}}}$, which is different in sign as comparing with that of the first two contributions.
} Hence, a conspiracy theory of cancellation is assumed to overcome the problem of too large resonance contributions to parameter $a$, or more accurately, $T(s)$ near $s=0$.
  In practice,  we therefore use the $\mathcal{O}(p^1)$ $\chi$PT result plus a polynomial background as the input $\mathrm{Im}_LT$, i.e.:
\begin{equation}\label{op1plusp}
	\mathrm{Im}_LT(s) = \mathrm{Im}_LT^{(1)}(s) + \mathrm{Im}_L[\dfrac{a+bs}{\sqrt{s}}]\ ,
\end{equation}	where $a$ and $b$ are simply two free parameters without relating to resonance parameters anymore.
The fit   gives  {$N(s_0)=0.57$}, $a=-2.39$ GeV and $b=-6.27$ GeV$^{-1}$, and one second sheet pole is found with
$\sqrt{s} = 0.95- 0.25i$ GeV without sizable spurious branch cut contributions. }

Since the mass of the $\rho$ meson is fixed we also tried the case of $\mathcal{O}(p^1)$ $\chi$PT results plus the $\rho$ meson exchange term  and a polynomial. That is,
\begin{equation}\label{rhopoly}
\mathrm{disc}\,T(s) = \mathrm{disc}\,T^{(1)}(s) +\mathrm{disc}\,T^{\rho}(s)+ \mathrm{d\mathrm{\mathrm{}}isc}\,[\dfrac{a+bs}{\sqrt{s}}]\ .
\end{equation}	
In this case the $\rho$ meson exchange produces an extra arc in $s$ plane~\cite{Ma:2020sym}, see Fig.~\ref{niuye}.
\begin{figure}[h]
\begin{center}
    \includegraphics[width=10cm]{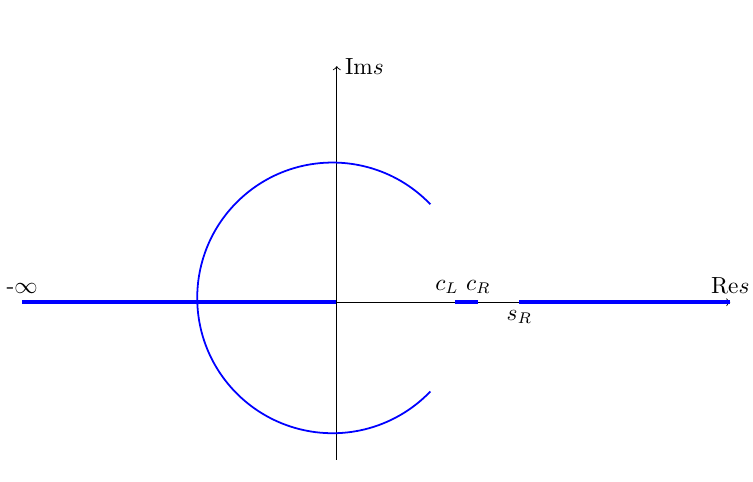}
    \caption{ The $l.h.c.$ caused by $t$-channel $\rho$ meson exchange
(circular arc~\cite{Ma:2020sym}); $u$-channel exchange (line segment from $c_L$ to $c_R$ ). {The branch point $d$ satisfies $|d|=m_N^2-m_\pi^2$~\cite{Ma:2020sym}} }\label{niuye}
    \end{center}
\end{figure}
We get
${N(s_0)=0.61,}$ $a =-7.88$ GeV, $b=-8.00$ GeV$^{-1}$,
and one second sheet pole is found located at
\bqa
\sqrt{s} = 0.90- 0.20i \mathrm{GeV}\ .
\eqa
These solutions are not stable -- there exist other solutions but with similar behaviours. In all cases the contributions from the spurious branch cuts  are negligible  as comparing with the results from the ${\cal O}(p^2)$ case, and   the $N^*(890)$ pole location remains stable.
It is noticed that the contribution from the arc cut generated by $t$-channel $\rho$ exchange is very small, e.g., it only contributes 1.5$^\circ$ at $\sqrt{s}=1.16$ GeV.

The ``spectral" function in this case is plotted in Fig.~\ref{imF_meson}. Different contributions to the phase shift according to the PKU decomposition are plotted in Fig.~\ref{pku_meson}.
\begin{figure}[H]
	\centering
	\includegraphics[width=10cm]{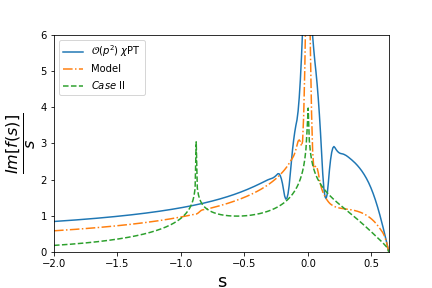}
	\caption{Comparison among different ``spectral" functions. {The singular behaviors of $T(s)$ at $s=0$ are $\mathcal{O}(s^{-5/2})$, $\mathcal{O}(s^{-1/2})$ and  $\mathcal{O}(s^{0})$ for $\mathcal{O}(p^2)$ $\chi$PT , model Eq.~(\ref{rhopoly}) and $Case$ II, respectively.}}\label{imF_meson}
\end{figure}
\begin{figure}[H]\centering
	\includegraphics[width=10cm]{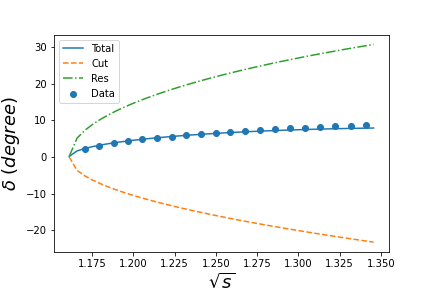}
	\caption{Fit results using Eq.~(\ref{rhopoly}). 
Phase shift decomposition: only contributions from physical ingredients are plotted including their summation `Total'. {It clearly demonstrates that  spurious contributions cancel each other,  otherwise curve `Total' cannot get close to the data.} }\label{pku_meson}
\end{figure}

{ Before  closing the discussions on numerical calculations we would like to stress that major physical outputs rely very little on the choice of the cutoff parameter $\Lambda_R^2=1.48$GeV$^2$. For example, setting $\Lambda_R^2=2.0,2.5$GeV$^2$ in model Eq.~(\ref{rhopoly}), lead to almost the same pole location at $\sqrt{s}=0.897-i0.193$GeV.}

In above,
we have made a rather long and exhaustive analyses which has to be stopped somewhere with some regrets. One is that
all the calculations made in this paper are performed at tree level only. At loop level, there are of course dynamical cut contributions, like the circular cut. The latter is estimated in Ref.~\cite{Wang:2018nwi} using the complete $\mathcal{O}(p^3)$ $\chi$PT input and it is found that {the sign of the circular cut contribution may vary depending on the choice of cutoff parameter, but always remains small in magnitude: e.g., at $\sqrt{s}=1.16$ GeV its contribution to the phase shift is 0.2$^\circ$  when $s_c=0.32$ GeV$^2$, and -1.7$^\circ$ when $s_c=-0.08$ GeV$^2$.} Since when evaluating the circle there is no problem like what happens at $s=0$, we think this estimation on the smallness of the circular cut contribution is at least qualitatively reasonable. See Fig.~\ref{cicular}, the cutoff parameter $s_c=0.32$GeV$^2$ corresponds to evaluating the $l.h.c.$ region covered by the green dashed circle, which can be estimated by chiral perturbation theory; whereas $s_c=-0.08$GeV$^2$ corresponds to that covered by the red dotted circle, which is required by best fit. The estimations made in Refs.~\cite{Wang:2018nwi}\cite{Wang:2017agd} pointed out that the region where $\chi$PT calculation can be safely used is not enough to generate the $N^*(890)$,\footnote{More precisely, the pole position is not stable. For example when taking $s_c=0.32$ GeV$^2$ the $N^*(890)$ degrades into two deep virtual poles.} i.e., certain help from the contribution in the region $s\in(-\infty, 0.32)$ is needed.\footnote{Taking $s_c=0.32$ GeV$^2$  will cause disasters in other channels as well \cite{Wang:2018nwi}.} The singularity in the ``spectral" function at $s=0$ seems to be helpful. It is realised that the rescue task is easily fulfilled by looking at Fig.~\ref{imF_meson}. The fit Case II only contains a weak singularity at $s=0$, i.e., $T(0)\propto \mathrm{const}$ while its contribution in the segment $(0.32, s_L]$ is much weaker than the $\mathcal{O}(p^2)$ ones, but it still affords a pole. The real situation should be much more optimistic.
\begin{figure}[h]
\begin{center}
    \includegraphics[width=11cm]{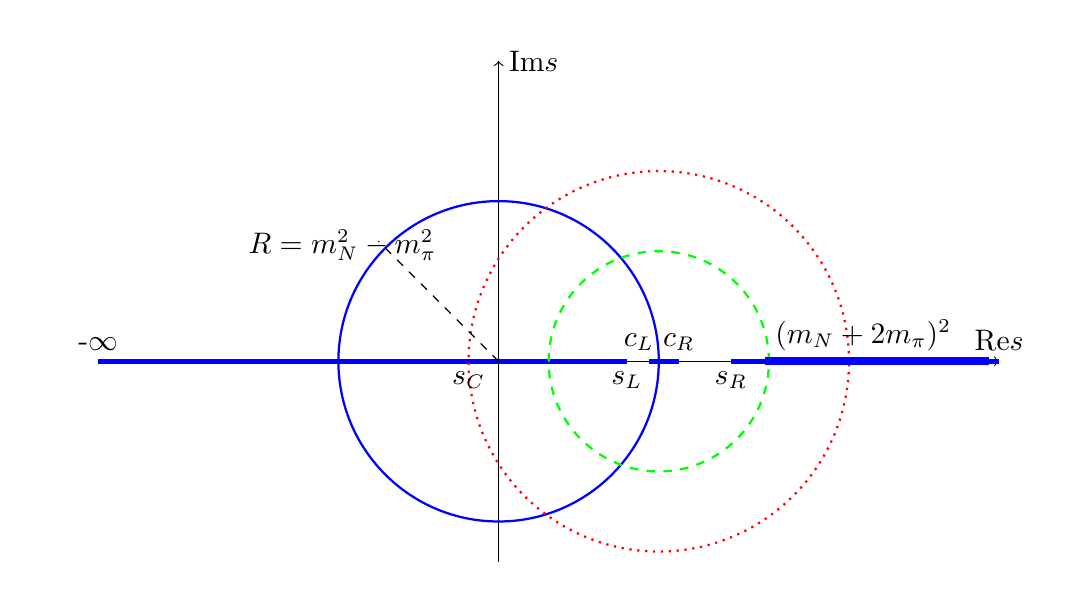}
    \caption{Region of $l.h.c.$ being used in Ref.~\cite{Wang:2018nwi}. The cutoff parameter $s_c=0.32$GeV$^2$ corresponds to evaluating the $l.h.c.$ region covered by the green dashed circle, which can be estimated by chiral perturbation theory; whereas $s_c=-0.08$GeV$^2$ corresponds to that covered by red dotted circle, which is required by best fit.  }\label{cicular}
    \end{center}
\end{figure}
It is noticed that the model Eq.~(\ref{op1plusp}) behaves quite like $\mathcal{O}(p^2)$ $\chi$PT results in the region $(0.32, s_L)$
and it is expected continuously to work in the region $(-\Lambda^2_L, -\epsilon)\cup(+\epsilon, 0.32)$, where $\Lambda^2_L$ is estimated to be around $R=m_N^2-m_\pi^2$ for example. We make a test by setting $\epsilon\simeq 0.05$ GeV$^2$ and cut off the peak around $s=0$ in the `spectral' function when $s\in(-\epsilon, +\epsilon)$. The $N^*(890)$ emerges stubbornly with a location {$\sqrt{s}=0.89-0.24i$ GeV.}

Hence, we conclude that the $l.h.c.$ contributions in total to the phase shift is sizable, based on which the $N^*(890)$ survives with  a rather stable pole location. Considering the level of accuracy of our calculations, we do not try to give statistical error bars here in this paper.

\section{Discussions and conclusions}

It may be somewhat amazing to claim something new in a field under extensive studies for more than half a century. However, according to our studies, a subthreshold broad resonance has well chance of being existed if the $s$-wave  phase shift steadily rise above the threshold as a convex curve. {The discussion made in the end of the last section suggests that the kinematical singularity structure  at $s=0$  plays a rather important role. {This is not surprising. One even finds examples in extreme cases that a pole can be generated totally for kinematic reasons.} For example, in $J=0,I=2$ channel of $\pi\pi$ scatterings, there exists a virtual pole which can be understood from pure kinematical reasons and it  brings important contributions to the phase shift~\cite{Zhou:2004ms}, and can be proved to exist rigorously~\cite{Guo:2007ff}.  Another example is the companionate virtual state of the nucleon, which can also be explained for pure kinematical reasons~\cite{Wang:2017agd}.    }

It is also interesting to notice that the $N^*(890)$ state  may be related to the lowest lying $1/2^-$ baryon states suggested by Azimov dated back in 1970~\cite{Azimov:1970ei}  named as $N^\prime$ there,  having been searched for desperately since then~\cite{Azimov:2003bb}. Contrary to the original proposal that the lowest lying nucleon counterpart lie above the $\pi N$ threshold, or at least lie above the nucleon mass (on the 1st sheet), the pole position named $N^*(890)$ determined in Refs.~\cite{Wang:2018nwi}\cite{Wang:2018gul}\cite{Wang:2017agd} as well as in this paper,  easily escapes of all bounds and limits set up previously~\cite{Azimov:2003bb}. The mass (width) difference may be explained by a familiar mechanism that when a strong coupling  gradually be turned on, the pole will move from the real axis above the threshold to the left of $s$ plane off the real axis. {Nevertheless, it may also be possible that the $N^*(890)$ `resonance'  be a virtual pole on $k$ plane, as already suggested in Ref.~\cite{Wang:2017agd}. In the latter situation, the `width' of $N^*(890)$ does not need to have any relations with particle decays.}  There still remains a lot of work to be done to identify  $N^\prime$ and $N^*(890)$. We plan to investigate this question in future.

 The $N/D$ calculations discussed in this paper are of rather complicated dynamical ones, however the production representation has been shown to be useful in providing us a simple and pictorial way of understanding the essence of the $N/D$ calculations: {in the language of this work}, the evidence on the existence of $N^*(890)$ seems to be  partly from the peculiar singularity structure of the background integral defined in Eqs.~(\ref{ffun}).
It reads that if $T(s=0)$ does not vanish (or does not vanish fast enough), then an $s$-wave subthreshold resonance exist, in the most attractive channels.\footnote{Similar observations are obtained, from a different point of view.~\cite{Xiao:2016dsx}.}  This may even be a rather universal phenomenon, if the background contributions are universally negative, as suggested by quantum scattering theory~\cite{Regge:1958ft} and repeatedly verified by calculations in quantum field theories.~\cite{Xiao:2000kx} \cite{Zheng:2003rw}\cite{Zhou:2004ms}

It is apparent that the existence of a light $1/2^-$ nucleon state is crucial for the completion and establishment of the lowest lying $1/2^-$ octet baryons as suggested in Ref.~\cite{Azimov:1970ei}, if ever it exists\footnote{For more information on the status of the octet baryons, see the talk of Igor Strakovsky given at $EHS-2019$, York, UK, December 2019.}. It will definitely improve our understanding of  strong interaction physics as well. For example, the $N^*(890)$ state, if exists, will definitely force us to rethink the possible physics behind $f_0(500)$ and $K(700)$. Another question it raises is how to interpret  spontaneous chiral symmetry breaking  more properly. The textbook explanation on this point is that the axial charge $Q_A$ commutes with the strong interaction hamiltonian, hence if chiral symmetry were not broken parity doublets would appear in nature. But what was observed as the lowest lying $1/2^-$ nucleon is $N^*(1535)$, the non-degeneracy of its mass comparing with the nucleon mass therefore indicates that chiral symmetry is spontaneously broken. The emergence of $N^*(890)$ may also bring new thinking on the related physics.

{Finally, it is also pointed out in this paper, that there exist two virtual poles located on the real axis, outside but very close to the $u$ channel cut $[c_L, c_R]$. Their existence is proven relying on the validity of chiral expansions  up to all orders.}

The authors would like to thank Zhi-hui Guo and De-Liang Yao for helpful discussions at various stage of this work, {and Ulf-G.~Mei{\ss}ner for a careful reading of the manuscript and useful suggestions}. Especially we thank Igor Strakovsky at George Washington University, for very interesting information on the $1/2^-$ octet baryons. This work is support in part by National Nature Science Foundations
of China under contract number 11975028 and 10925522, and by the {Deutsche Forschungsgemeinschaft (DFG, German Research Foundation) - Project-ID 196253076 -TRR 110}.

\renewcommand\refname{Reference}
\bibliographystyle{h-physrev}
\bibliography{NoverD}

\begin{thebibliography}{10}

\bibitem{Wang:2018nwi}
Y.~F. Wang, D.~L. Yao, and H.~Q. Zheng,
\newblock Chin. Phys. C {\bf 43}, 064110 (2019).

\bibitem{Wang:2018gul}
Y.~F. Wang, D.~L. Yao, and H.~Q. Zheng,
\newblock Front. Phys. (Beijing) {\bf 14}, 24501 (2019).

\bibitem{Wang:2017agd}
Y.~F. Wang, D.~L. Yao, and H.~Q. Zheng,
\newblock Eur. Phys. J. C {\bf 78}, 543 (2018).

\bibitem{Zheng:2003rw}
H.~Q. Zheng {\em et~al.},
\newblock Nucl. Phys. A {\bf 733}, 235 (2004).

\bibitem{Zhou:2006wm}
Z.~Y. Zhou and H.~Q. Zheng,
\newblock Nucl. Phys. A {\bf 775}, 212 (2006).

\bibitem{Zhou:2004ms}
Z.~Y. Zhou {\em et~al.},
\newblock JHEP {\bf 02}, 043 (2005).

\bibitem{Xiao:2000kx}
Z.~G. Xiao and H.~Q. Zheng,
\newblock Nucl. Phys. A {\bf 695}, 273 (2001).

\bibitem{He:2002ut}
J.~Y. He, Z.~G. Xiao, and H.~Q. Zheng,
\newblock Phys. Lett. B {\bf 536}, 59 (2002),
\newblock [Erratum: Phys. Lett. B 549, 362--363 (2002)].

\bibitem{Ma:2020sym}
Y.~Ma, W.~Q. Niu, Y.~F. Wang, and H.~Q. Zheng,
\newblock Commun. Theor. Phys. {\bf 72}, 105203 (2020).

\bibitem{Ma:2020hpe}
Y.~Ma, W.~Q. Niu, D.~L. Yao, and H.~Q. Zheng,
\newblock Chin. Phys. C {\bf 45}, 014104 (2021).

\bibitem{Cao:2021kvs}
X.~H. Cao, Y.~Ma, and H.~Q. Zheng,
\newblock Phys. Rev. D {\bf 103}, 114007 (2021).

\bibitem{Yao:2020bxx}
D.~L. Yao, L.~Y. Dai, H.~Q. Zheng, and Z.~Y. Zhou,
\newblock Rep. Prog. Phys. {\bf 84}, 076201 (2021).

\bibitem{Gasparyan:2010xz}
A.~Gasparyan and M.~F.~M. Lutz,
\newblock Nucl. Phys. A {\bf 848}, 126 (2010).

\bibitem{Kennedy:1961}
J.~Kennedy and T.~D. Spearman,
\newblock Phys. Rev. {\bf 126}, 1596 (1961).

\bibitem{Hoferichter_2016}
M.~Hoferichter, J.~Ruiz~de Elvira, B.~Kubis, and U.-G. Mei{\ss}ner,
\newblock Physics Reports {\bf 625}, 1 (2016).

\bibitem{Alarcon:2012kn}
J.~M. Alarcon, J.~Martin~Camalich, and J.~A. Oller,
\newblock Annals Phys. {\bf 336}, 413 (2013).

\bibitem{Chen:2012nx}
Y.~H. Chen, D.~L. Yao, and H.~Q. Zheng,
\newblock Phys. Rev. D {\bf 87}, 054019 (2013).

\bibitem{Bruns:2010sv}
P.~C. Bruns, M.~Mai, and U.~G. Mei{\ss}ner,
\newblock Phys. Lett. B {\bf 697}, 254 (2011).

\bibitem{Siemens:2016hdi}
D.~Siemens {\em et~al.},
\newblock Phys. Rev. C {\bf 94}, 014620 (2016).

\bibitem{Siemens:2016jwj}
D.~Siemens {\em et~al.},
\newblock Phys. Lett. B {\bf 770}, 27 (2017).

\bibitem{Siemens:2017opr}
D.~Siemens {\em et~al.},
\newblock Phys. Rev. C {\bf 96}, 055205 (2017).

\bibitem{Jakob:1969hn}
H.~P. Jakob and F.~Steiner,
\newblock Z. Phys. {\bf 228}, 353 (1969).

\bibitem{Hoferichter:2015tha}
M.~Hoferichter, J.~Ruiz~de Elvira, B.~Kubis, and U.-G. Mei{\ss}ner,
\newblock Phys. Rev. Lett. {\bf 115}, 192301 (2015).

\bibitem{Yao:2016vbz}
D.~L. Yao {\em et~al.},
\newblock JHEP {\bf 05}, 038 (2016).

\bibitem{Scherer:2012xha}

\newblock S.~Scherer and M.~R. Schindler{\em , A primer for chiral perturbation
  theory,} Vol. 830 (Springer Science \& Business Media, 2011).

\bibitem{Sanz-Cillero:2013ipa}
J.~J. Sanz-Cillero, D.~L. Yao, and H.~Q. Zheng,
\newblock Eur. Phys. J. C {\bf 74}, 2763 (2014).

\bibitem{Qin:2002hk}
G.~Y. Qin, W.~Z. Deng, Z.~Xiao, and H.~Q. Zheng,
\newblock Phys. Lett. B {\bf 542}, 89 (2002).

\bibitem{Ecker:1988te}
G.~Ecker, J.~Gasser, A.~Pich, and E.~de~Rafael,
\newblock Nucl. Phys. B {\bf 321}, 311 (1989).

\bibitem{Bernard:1996gq}
V.~Bernard, N.~Kaiser, and U.-G. Meissner,
\newblock Nucl. Phys. A {\bf 615}, 483 (1997).

\bibitem{Appelquist:1974tg}
T.~Appelquist and J.~Carazzone,
\newblock Phys. Rev. D {\bf 11}, 2856 (1975).

\bibitem{Ovrut:1980eq}
B.~A. Ovrut and H.~J. Schnitzer,
\newblock Phys. Rev. D {\bf 22}, 2518 (1980).

\bibitem{Guo:2007ff}
Z.~H. Guo, J.~J. Sanz-Cillero, and H.~Q. Zheng,
\newblock JHEP {\bf 06}, 030 (2007).

\bibitem{Azimov:1970ei}
Y.~I. Azimov,
\newblock Phys. Lett. B {\bf 32}, 499 (1970).

\bibitem{Azimov:2003bb}
Y.~I. Azimov, R.~A. Arndt, I.~I. Strakovsky, and R.~L. Workman,
\newblock Phys. Rev. C {\bf 68}, 045204 (2003).

\bibitem{Xiao:2016dsx}
Z.~Xiao and Z.-Y. Zhou,
\newblock Phys. Rev. D {\bf 94}, 076006 (2016).

\bibitem{Regge:1958ft}
T.~Regge,
\newblock Nuovo Cim. {\bf 8}, 671 (1958).

\end{thebibliography}
\begin{newpage}
\begin{appendix}
\section{Phenomena of the cut $[c_L,c_R]$ induced by $u$ channel nucleon exchanges }\label{appendix}
The $u$ channel nucleon pole exchange diagram will contribute to the partial wave amplitude a cut $\in[c_L,c_R]$, with $c_L=\frac{(m_N^2-m_\pi^2)^2}{m_N^2}$ and $c_R=m_N^2+2m_\pi^2$. The point $s=c_L$ is reached when $z_s=-1$, $u=m_N^2$ and $s=c_R$ is reached when  $z_s=+1$, $u=m_N^2$ (see Eq.~(\ref{usz})). More precisely, at branch points $c_L$ and $c_R$ one gets the leading term of the partial wave amplitude:
\bqa\label{a1}
 {s\to {c_L}:}&&\,\,\,T(s)\to -\frac{g^2 m_N^4 }{16 \pi  F^2 \left(4 m_N^2-m_\pi^2\right)}\ln \frac{s-c_L}{c_L-c_R}\ ,\nonumber\\
 {s\to {c_R}:}&&\,\,\,T(s)\to\frac{g^2 m_N^2 (m_N^2+2 m_\pi^2)} {\pi  F^2 (4 m_N^2-m_\pi^2)}\ln \frac{c_R-c_L}{s-c_R}\ ,
\eqa
which are solely from the $u$ channel nucleon pole term, and are hence exact, i.e., receiving no chiral corrections.
Based on Eq.~(\ref{a1}) one further finds
\begin{equation}\label{SCLCR}
\begin{split}
& s\to c_L, \quad	S^{}\simeq A_{c_L}+B_{c_L}\ln\frac{s-c_L}{c_L-c_R}\ ,\\
	&s\to c_R, \quad	S^{}\simeq A_{c_R}+B_{c_R}\ln\frac{s-c_R}{c_R-c_L}\ ,
\end{split}
\end{equation}
in which the coefficients read
\begin{equation}
	\begin{split}
	&A_{c_L}=A_{c_R}=1+\frac{g^2 m_N m_\pi}{8 \pi  F^2}+O\left(m_\pi^3\right)\ ,\\
		&B_{c_L}=B_{c_R}=\frac{g^2 m_N m_\pi}{16 \pi  F^2}+O\left(m_\pi^3\right)\ .
	\end{split}
\end{equation}
These equations are obtained from Born term calculations,
$A_{c_L}$ ($B_{c_L}$) and $A_{c_R}$ ($B_{c_R}$) differ at $O(m_\pi^3)$ level. It is worth emphasizing that
$A_{c_L}$ and $A_{c_R}$ may receive chiral corrections, {but $B_{c_L}$ and $B_{c_R}$ do not,} since the latter is related to the residue of the $u$ channel nucleon pole.

One important conclusion one can draw is that $S(c_L), S(c_R)\to -\infty$ which are exact (correct at least to any order of chiral expansions) and are  immune of any loop corrections. Remember that $S=+1$ at $s_L$ and $s_R$ by definition, and $S(s)$ is real when $s\in (s_L, c_L)\cup(c_R, s_R)$,  one finds that there have to be two $S$ matrix zeros: one below $c_L$ and another above $c_R$, on the real axis.\footnote{In a $K$-matrix unitarization,  the $S$ matrix  no longer diverges at $s=c_L, c_R$, the two virtual poles however still exist, and no major conclusions change.} Their locations (denoted by $v_L$ and $v_R$) are:
\begin{equation}\label{VS}
	\begin{split}
	& v_L=c_L-(c_R-c_L)e^{-A_{c_L}/B_{c_L}}\ ,\\
	&v_R=c_R+(c_R-c_L)e^{-A_{c_R}/B_{c_R}}\ .
	\end{split}
\end{equation}
Even more surprisingly, these two virtual poles in total give a large contribution to the phase shift. E.g., they give
roughly $50^\circ$ at $\sqrt{s}=1.16$GeV, which seems to completely destroy the picture presented in Figs.~\ref{ndphase} and \ref{pku_meson}. The solution to this apparent paradox is rather tricky, which  we discuss in the following.

In the derivation of the production representation, Eqs.~(\ref{ffun}) and (\ref{ffun'}), it is generally assumed  that the branch cut singularity structure of $S^{{cut}}$ ($=\exp\{2i\rho(s)f(s)\}$) and $f(s)$ (or $\ln{S^{cut}}$) being the same. This is not always being true: when $S^{cut}$ is real and negative at certain point, $\ln S^{cut}$ will have to be discontinuous when  the sign of the imaginary part of $S^{cut}$ changes. This situation indeed happens in the present situation. {Recall that at $O(p^1)$ level
 \begin{equation}\label{op1}
\begin{split}
T^{(1)}=&
-\frac{g^2 \left(-2 m_N^2 m_\pi^2-m_N^2 s-m_\pi^2 s+s^2\right)}{32 \pi  F^2 \left(s-m_N^2\right)}+\frac{-m_N^2-m_\pi^2+s}{32 \pi  F^2}\\
&+\frac{m_N \left(-m_N^2+m_\pi^2+s\right)}{32 \pi  F^2\sqrt{s}}-\frac{g^2 m_N \left(-m_N^4+m_N^2 m_\pi^2+m_N^2 s+2 m_\pi^2 s\right)}{32 \pi  F^2 \left(s-m_N^2\right)\sqrt{s}}\\
& +\frac{g^2 m_N^2 s^2 \left(-m_N^2-m_\pi^2+s\right)}{16 \pi  F^2 \left(s-c_L\right)^2 \left(s-c_R\right)^2} \left(\frac{m_N^2}{s}\left(s-c_L\right) \left(\log \frac{s-c_L}{s-c_R}+\log \frac{m_N^2}{s}\right)-s\rho(s)^2\right)\\
&-\frac{g^2 m_N^3 s^2 \left(-m_N^2+m_\pi^2+s\right)}{16 \pi  F^2 \left(s-c_L\right)^2 \left(s-c_R\right)^2\sqrt{s}} \left(\left(s-c_R\right) \left(\log \frac{s-c_L}{s-c_R}+\log \frac{m_N^2}{s}\right)-s\rho(s)^2\right)\ .
\end{split}
\end{equation}
It is seen from the above expressions that the cut $L\in (-\infty,0)\cup(c_L,c_R)$. There are two sources contributing to the imaginary part of $T(s)$ when $s\in(-\infty,0)$: one comes from the kinematical $\sqrt{s}$ while another comes from the logarithmic function; when $s\in(c_L,c_R)$ $\mathrm{Im}T$ solely comes from the $u$ channel nucleon pole exchange. On $(c_L,c_R)$ the imaginary part of $T^{(1)}$ reads:
\begin{eqnarray}\label{IMT1}
 &&\mathrm{Im}T^{(1)}(s) =
 \frac{g^2 m_N^2 s^2 }{16F^2 \left(s-c_L\right)^2 \left(s-c_R\right)^2} \nonumber\\
  && \times\left[\dfrac{m_N^2}{s}(s-c_L)\left(-m_N^2-m_\pi^2+s\right)-\dfrac{m_N}{\sqrt{s}}(s-c_R)\left(-m_N^2+m_\pi^2+s\right)\right]\ ,
\end{eqnarray}
from which it is seen that the $\mathrm{Im}T^{(1)}$ develops a zero at $s=s_c\simeq  m_N^2-\frac{m_\pi^4}{2m_N^2}$ and changes sign when $s$ crosses $s_c$. It is important to realize that Eq.~(\ref{IMT1}) is immune of chiral perturbation corrections.}

\begin{figure}[H]
	\includegraphics[width=7cm]{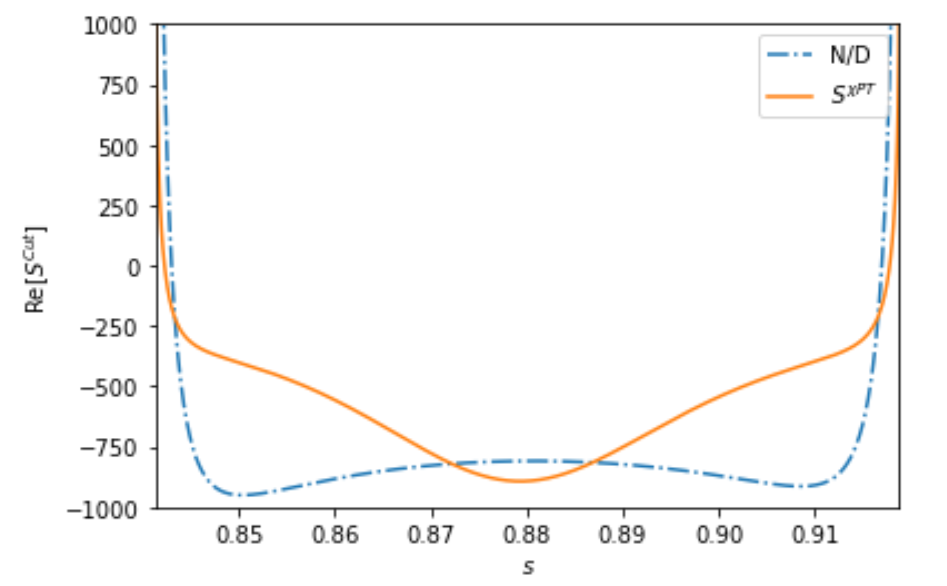}
	\includegraphics[width=7cm]{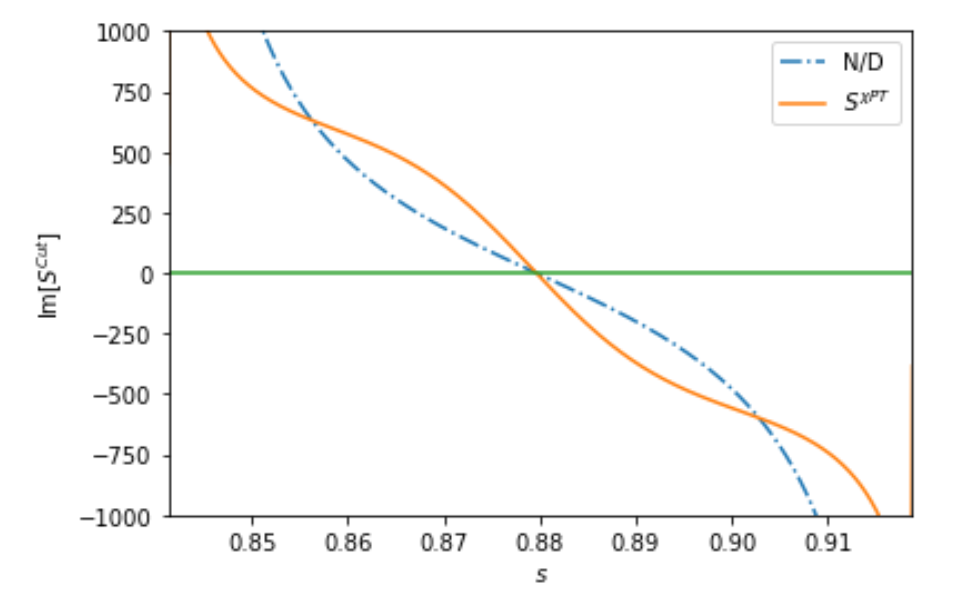}
	\caption{Real (left) and imaginary (right) part of $S^{cut}(s)$ when $s$ lies in $(c_L,c_R)$. The dot-dashed line comes from N/D solution of Eq.~(\ref{op1plusp}), the yellow solid line is obtained from $O(p^1)$ $\chi$PT results.}\label{scut}
\end{figure}
\end{appendix}
\end{newpage}
From Fig.~\ref{scut},
it is found that the imaginary part of $S^{cut}$ vanishes at one point $s_c$ which is close to $m_N^2$ meanwhile the real part of $S^{cut}$ is negative. In Fig.~\ref{scut} $S^{cut}$ is calculated by
\be\label{scutnew}
S^{cut}={S^{phys}\over\prod_p S^{p}\times S^{v_L}\times S^{v_R}}
\ee
where the newly found virtual poles are included, and $S^{phys}$ are approximated by  both the unitary amplitude obtained by fitting Eq.~(\ref{op1plusp}) and the pure $ O(p^1)$ perturbation amplitude.~\footnote{The `$O(p^1)$ calculation' in Fig.~\ref{scut} is done via Eq.~(\ref{scutnew}) in the following way: $S^{phys}$ is calculated in $O(p^1)$, `poles' are picked up by searching for zeros of $S^{phys}$, though they are actually not second sheet poles for lacking of unitarity.}
From Fig.~\ref{scut} it is seen that at $s=s_c$, $\ln S^{cut}$
has to develop a discontinuity and hence a branch cut emerges crossing $s_c$. It is numerically checked that the cut is an arc on the complex $s$-plane in $N/D$ solutions. It should be emphasized here that this unexpected additional cut does not pollute $S^{cut}$, since across the cut the discontinuity
of $\ln S^{cut}$ is $2i\pi$ and has no influence to the value of an exponential. Though not producing any trouble in the analyticity structure of $S^{cut}$, the additional cut of $\ln S^{cut}$, or the dispersive representation of function $f$ defined in Eq.~(\ref{ffun}) has to be changed since the integration contour has to be modified.

The problem found above is rather severe since the distorted contour may depend on numerics and hence being impossible to control. However, it can be overcome by the following consideration. Define
\begin{equation}\label{overlinef}
	\bar f(s) = \dfrac{\ln-S^{cut}}{2i\rho(s)}- \dfrac{\pi}{2\bar\rho(s)}
\end{equation}
where the   function $\bar\rho(s)$ is the `deformed' $\rho(s)$ with its cut $\in [s_L, s_R]$, while the cut of the latter is defined on $(-\infty, s_L]\cup [s_R,+\infty)$. Notice that $\bar \rho(s)$ and $\rho(s)$ are identical in the physical region.
Function $\bar f(s)$ is identical to $f(s)$ when $s$ in the physical region as well but differs in cut alignment.
Particularly $\bar f$ no longer contains the arc cut of $f$, since $s_c$ is not a branch point of $\bar f$ anymore. However, $\bar f$ contains an additional cut induced by $\bar\rho$ which is absent in $f$. Both the two terms on the $r.h.s.$ of Eq.~(\ref{overlinef}) contain cuts on $[s_L,s_R]$, but the two cuts cancel each other when {$s\in [c_R,s_R]$}. Hence the left cut of $\bar f$ on the real axis is actually $[-\infty, c_R]$, comparing with the left cut of $f$ on the real axis: $[-\infty, s_L]\cup[c_L,c_R]$. The dispersive integral representation of $\bar f$ can be written as:
\begin{equation}\label{fps}
	\begin{split}
	\bar f(s) = &\dfrac{s}{\pi}\int_L\dfrac{\mathrm{Im}[\ln {-S^{cut}(s^\prime)}/(2i\rho(s^\prime))]}{s^\prime(s^\prime-s)}ds^\prime\\
&{+\dfrac{s}{\pi}\int_{s_L}^{c_L}\dfrac{\mathrm{Im}[\ln {-S^{cut}(s^\prime)}/(2i\rho(s^\prime))]}{s^\prime(s^\prime-s)}ds^\prime}\\
	&{-s\int_{s_L}^{c_R}\dfrac{\mathrm{Im}[1/(2\bar\rho(s^\prime))]}{s^\prime(s^\prime-s)}ds^\prime}\ .
	\end{split}
\end{equation}
In the first term on the $r.h.s.$ of the above equation, the integration domain $L=(-\infty, s_L]\cup[c_L,c_R]$, i.e., the same as that in Eq.~(\ref{ffun}). Actually the first
term is identical to $f$ defined in Eq.~(\ref{ffun}) as in the integrand $S^{cut}$ can actually be replaced by $S^{phys}$. To prove this it is realized that, firstly, on $(-\infty, s_L]$, $\mathrm{Im}[\ln {-S^{cut}}/(2i\rho)]=-\ln {|S^{cut}|}/(2\rho)$, and $|S^{cut}|=|S^{phys}|$. Secondly, the integral on $[c_L,c_R]$ can be recast as
\begin{equation}
	\begin{split}
	&\dfrac{s}{\pi}\int_{c_L}^{c_R}\dfrac{\mathrm{Im}[\ln {-S^{cut}(s^\prime)}/(2i\rho(s^\prime))]}{s^\prime(s^\prime-s)}ds^\prime = \dfrac{s}{2i\pi}\int_{c_L}^{c_R}\dfrac{\ln[S^{cut}_+/S^{cut}_-]}{2i\rho(s^\prime)s^\prime(s^\prime-s) }ds^\prime\\
	&=\dfrac{s}{2i\pi}\int_{c_L}^{c_R}\dfrac{\ln[S^{phys}_+/S^{phys}_-]}{2i\rho(s^\prime)s^\prime(s^\prime-s) }ds^\prime
	 =	\dfrac{s}{\pi}\int_{c_L}^{c_R}\dfrac{\mathrm{Im}[\ln {S^{phys}(s^\prime)}/(2i\rho(s^\prime))]}{s^\prime(s^\prime-s)}ds^\prime\ ,
	\end{split}
\end{equation}
where $S_{\pm}\equiv S(s\pm i\epsilon)$. Hence we complete the proof.

The second integral in Eq.~(\ref{fps}) is analytically integrable once it is realized that $S^{cut}$ is positive definite along $[s_L, c_L]$, and the third integral is also integrable. {The sum of the two integrals gives a contribution to the phase shift exactly canceled by the contribution from the two virtual poles, when their positions are taken at $c_L$ and $c_R$.} If the two virtual pole locations are fixed by Eq.~(\ref{VS}), the net effects of the sum of virtual poles and the  additional cut are vanishingly small (e.g., of order of ${10^{-3}}$ degrees at $\sqrt{s}=1.16$GeV).
{Therefore, The calculation using Eq.~(\ref{ffun}) is still {valid} with high accuracy, when one ignores the existence of the two virtual poles.} {Further, it is worth emphasizing that after such a surgery, there no longer exists the unwanted cut in $\ln S^{cut}$ crossing $s_c$, as checked by numerical analyses.} {Moreover, it may be worth pointing out that in numerical analyses there may appear additional cuts in  $\ln S^{cut}$ as well on the $s$ plane in distance, which is caused by the peculiar analyticity property of the logarithmic function. But it is easy to prove that it is not hazardous and can be simply ignored.}

The discussions made in this appendix can be extended to higher partial waves as well and some very interesting results appear, which will be presented elsewhere.
\end{document}